\documentclass[10pt]{article}
\usepackage[latin1,utf8]{inputenc}
\usepackage{graphicx}
\usepackage{epstopdf}
\usepackage{float} 
\usepackage[small,bf]{caption} 
\usepackage{amsmath}
\usepackage{url}
\usepackage{mathpazo}

\title{Overground gait transitions are not sharp but involve gradually changing walk-run mixtures even over long distances}

\author{
Nicholas S. Baker$^{1}$, Leroy L. Long III$^2$, and Manoj Srinivasan$^{3}$ \\
$^{1,3}$Mechanical and Aerospace Engineering, The Ohio State University, OH 43210 \\
$^{2}$Mechanical Engineering Technology, Sinclair Community College, OH 45402 \\
$^{3}$Program in Biophysics, The Ohio State University, OH 43210 \\
$^{3}$Email: \texttt{srinivasan.88@osu.edu}}
\date{}

\begin{document}
\maketitle

\begin{abstract}
Humans typically walk at low speeds and run at higher speeds. Previous studies of transitions between walking and running were mostly on treadmills, but real-world locomotion allows more flexibility. Here, we study overground locomotion over long distances (800 m or 2400 m) under time constraints, simulating everyday scenarios like going to an appointment. Unlike on treadmills, participants can vary both speed and gait during this task. We find that gait transition in this overground task occurs over a broad `gait transition regime' spanning average speeds from 1.9 m/s to 3.0 m/s. In this regime, people use mixtures of walking and running: mostly walking at low average speeds (around 1.9 m/s) and mostly running at high average speeds (3.0 m/s); the walk vs run fraction gradually changes between these speed limits. Within any walk-run mixture, there is a speed gap between the walking and running. These gait mixtures and their specific structure are predicted by energy optimality.  These findings extend earlier results from much shorter distance tasks, showing that similar energetic principles govern longer, more physically and cognitively demanding tasks. Overall, our results highlight the role of whole-task energy minimization including transients in shaping human locomotion.
\end{abstract}

\maketitle

\section{Introduction}
\begin{figure}[htbp]
\begin{center}
	\includegraphics{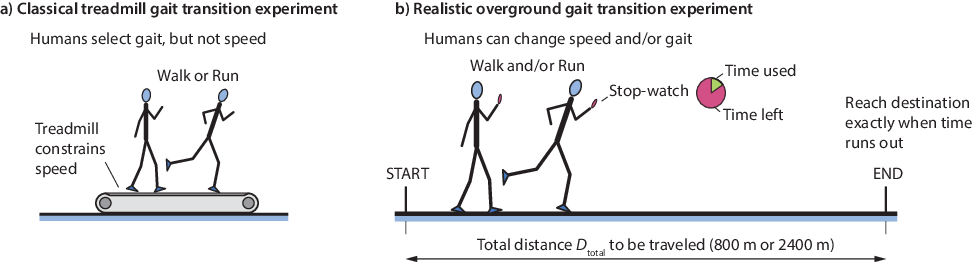}
\caption{\textbf{Experimental protocol.} a) The classic treadmill-bound gait transition used in prior studies experiment does not provide any instantaneous choice in speed to the human, constraining speed at every moment, so that participants are just asked to choose the gait they prefer at a given speed. b) Overground gait transition experiment used in the current study. Human participants are asked to cover a given distance $D_\mathrm{total}$ in a fixed amount of time $T_\mathrm{total}$, as if they have to travel on foot to make an appointment, exactly arriving on time. Participants carry a stop-watch counting down time remaining to prescribed duration and are free to choose and change their gait or speed whenever during each trial. Trials are repeated for different prescribed durations so as to achieve different average speeds.}
\label{fig:Fig1Expository}
\end{center}
\end{figure}

Humans and many other terrestrial animals exhibit a number of different gaits \cite{LongSrinivasan2013,Sri11,hildebrand1976analysis,collins1993coupled,Mar76}. Humans walk, run, and much more occasionally, skip \cite{minetti1998biomechanics,ackermann2012predictive}. Horses walk, trot, canter, and gallop, and more occasionally, use other gaits \cite{hildebrand1976analysis,Hoyt81,Rui05}. Transitions between such gaits have most commonly been studied using treadmills. In these treadmill gait transition experiments  (Figure \ref{fig:Fig1Expository}), the treadmill speed is changed slowly, either in a continuous fashion with some fixed acceleration, or in a series of acceleration phases alternating with constant speed phases \cite{Hre93,Hrel07a,Hrel07b,Dan03,VanCae10a,Seg07,Ray02,Tur99,Die95}. These experiments found that people switch between walking and running around 2 m/s, but sometimes, the walk to run transition speed was different from --- and higher than ---  the run to walk transition speed \cite{Die95,Hrel07a,Hrel07b}. Importantly, the transition speeds were different from that predicted by energy optimality \cite{Hre93,brill2021does}, even though there is extensive evidence indicating energetics as a determinant of locomotor choice in numerous other settings \cite{brown2021unified,seethapathi2015metabolic,Sri09,kuo2005energetic,maxwell2001mechanical,joshi2015walking,alexander1999energy,Sri11,Sri06,kuo2005energetic,Ale89,falisse2019rapid}. In such treadmill experiments, the gait transition is `sharp', that is, happens at a particular speed where there is a preference of running over walking, or vice versa. Of course, humans do not spend their lives on treadmills, so their behavior may not already be energy optimal for such gait transition tasks without considerable learning and experience near the gait transition speeds \cite{seethapathi2024exploration,selinger2019humans,choi2007adaptation}. Here, in contrast to these treadmill gait transition experiments, we show that overground gait transitions in realistic overground locomotion is more gradual and provide some potential clues for why there might exist distinct walk-to-run and run-to-walk speeds on a treadmill.

Say you need to travel on foot from your home to an important appointment a kilometer away at a particular time, which provides a time-deadline (Figure \ref{fig:Fig1Expository}). Unlike on a treadmill, where the speed is constrained, in this overground experiment, you can change speed or change gait.  If you start very early and have plenty of time, you might prefer to walk all the way. If you have very little time, you might need to run all the way. But if you had an intermediate amount of time, what might you do? Here, we perform this experiment for two long distances, 800 m and 2400 m, and show that humans systematically use a mixture of walking and running when there is an intermediate amount of time. That is, we show that for such overground tasks, there is not a sharp gait transition speed below which walking is preferred and above which running is preferred. Having this mixture of walking and running instead of a sharp gait transition speed is energy optimal \cite{LongSrinivasan2013}, and was earlier observed in humans by Long and Srinivasan \cite{LongSrinivasan2013} in a much shorter distance version of this task. So, the current manuscript can be considered an expansion of this previous study \cite{LongSrinivasan2013}, with the experimental contribution of the current study being consideration of much longer distances, but with additional contributions in terms of data analyses and computational work not present in the previous study \cite{LongSrinivasan2013}. Longer distances make the energy costs more salient to the human, while also making the task of adjusting speed and gait in an optimal manner more cognitively demanding. So, it is \textit{a priori} unclear whether humans will use walk-run mixtures as they did over short distances. This new data over longer distances also allows us to characterize how the preferred walking and running speeds, the fractions of walking and running, and the within-trial gait transition speed vary as a function of the mean locomotion speed, not previously characterized. We then show that these trends in the human walk-run mixtures are also qualitatively consistent with energy minimization. 

\section{Methods}
\paragraph{Experimental methods.} Experiments were approved by the Ohio State Institutional Review Board and human participants took part in them with informed consent. We had $N = 14$ participants (11 male, 3 female) with age 21.3 $\pm$ 2.9\,years (mean $\pm$ s.d.), mass 70.35 $\pm$ 13.99\,kg, height 1.80 $\pm$ 0.10\,m, and leg length 0.91 $\pm$ 0.26\,m. participants were instructed to travel one of two distances on foot, $D_\mathrm{total} =$ 800\,m or 2400\,m, in a pre-determined amount of time, $T_\mathrm{total}$, arriving exactly as allowed time expired, arriving neither early nor late (Figure \ref{fig:Fig1Expository}b and Table \ref{table:protocol}).  A stop-watch that counted down from $T_\mathrm{total}$ to zero was provided to the participant for reference throughout the trial. The $T_\mathrm{total}$ was not revealed to the participants until just moments before starting a trial to eliminate as much prior planning as possible. The participants received no additional instructions. The total distance and time constraints constrained only the average speed $v_\mathrm{avg} = D_\mathrm{total}/T_\mathrm{total}$, but not the instantaneous speeds throughout the trial, in contrast to a classic treadmill gait transition experiment (Figure \ref{fig:Fig1Expository}a). 

For each total distance $D_\mathrm{total}$, we used four different $T_\mathrm{total}$ corresponding to four different average speeds: 1.92, 2.24, 2.62, and 2.98 \,ms$^{-1}$. These speeds were chosen to be roughly in the treadmill gait transition regime \cite{Mar76,Thor87,Hre93,Hrel07a}, as well as the regime in which humans used a mixture of walking and running in our earlier study where we examined this protocol for a shorter distance \cite{LongSrinivasan2013}. Because of the large distances involved here, each participant either performed two trials both at the 2400\,m distance, or three trials, two at 800\,m and one at 2400\,m. In all, we performed $N_\mathrm{trial} =$ 42 total trials across the 14 participants, with at least 9 trials per average speed and 10.5 trials per speed on average. The trial assignment and trial order were randomized for each participant and across participants. The total duration was not revealed to the participants until a few moments before starting to eliminate as much pace planning as possible. All trials were performed on a 400\,m outdoor track. Locomotion speeds were measured using GPS (10\,Hz, VBOX Mini, Racelogic, mounted in a waist pouch) and was available for all but one subject for whom this data was corrupted. Trials were video recorded and used to determine durations of walking and running during each trial, as visually classified by the investigators. The investigators classified the trial as running if there was a flight phase or if the hip moved downward and upward on a compliant-looking leg during stance phase, even if there did not seem to be a flight phase (jogging or grounded running); walking bouts are the complements of these running bouts \cite{daley2018scaling,Bli93}.

\paragraph{Computational model.} Our energy-based account of these overground gait transitions is based on minimizing the energy cost of the entire locomotor task, including the transients like acceleration, deceleration, and gait transitions --- as opposed to minimizing something instantaneous or steady-state like the cost of transport \cite{Tuck75,Mar76,Sri09}. Earlier work had shown that this whole-task energy optimality framing explains locomotor phenomena better \cite{LongSrinivasan2013,seethapathi2015metabolic,brown2021unified}, with some recent work finding additional evidence \cite{carlisle2023optimization}.

For each trial, given travel distance $D_\mathrm{total}$ and time duration $T_\mathrm{total}$, we compute the energy optimal walking speed $V_\mathrm{walk}$ and running speeds $V_\mathrm{walk}$, and the respecting walking and running durations ($T_\mathrm{walk}$ and  $T_\mathrm{run}$). The individual gait durations must satisfy the total duration constraint $T_\mathrm{walk}+T_\mathrm{run} = T_\mathrm{total}$, and along with the speeds, they satisfy the total distance constraint $V_\mathrm{walk}T_\mathrm{walk} + V_\mathrm{walk}T_\mathrm{walk} = D_\mathrm{total}$. The total energy cost of a bout is computed as the sum of the walking energy cost, running bout energy cost, and the cost for the transitions between gaits as well as the changes in speed \cite{seethapathi2015metabolic,Ush03,LongSrinivasan2013}. The metabolic rate for walking and running as a function of speed are of the form $\dot{E} = a_0 + a_1 V + a_2 V^2$, with different coefficients \cite{LongSrinivasan2013,Steu09,Bob60}, the cost for changing speed is proportional to the kinetic energy change \cite{seethapathi2015metabolic}, and the gait change cost is equal to that in \cite{Ush03}.

These calculations are similar to that in \cite{LongSrinivasan2013} specialized to the task conditions herein, and are performed using the nonlinear optimization procedure fmincon in MATLAB; see \textit{Code availability} for links to the MATLAB code. Here, in addition, noting that the energy landscape tends to have relatively flat minima \cite{seethapathi2015metabolic,handford2014sideways,brown2021unified,selinger2019humans}, we provide a sensitivity analysis: we compute the set of all speeds and gait durations that are within 1\% and 2\% of the previously computed optimal energy cost. To do this, we scan over a range of walking time durations for each speed, and solve for the optimal walking and running speeds for each such walking duration along with the corresponding energy costs, which then allows us to characterize the speeds and durations within 1\% or 2\% from the optimal energy cost.

\begin{table}
\begin{center}
\begin{tabular}{|| l | c c c c c c c c||} 
 \hline
Trial ID & 1 & 2 & 3 & 4 & 5 & 6 & 7 & 8 \\ 
 \hline
 Distance $D_\mathrm{total}$ (m) & 800 & 800  & 800 & 800 & 2400 & 2400 & 2400 & 2400 \\
Duration $T_\mathrm{total}$ (min)	 & 14.00 & 12.00  & 10.25	 & 9.00  & 14.00  & 12.00  & 10.25  & 9.00 \\
Average speed $V_\mathrm{avg}$ (m/s) & 1.92 & 2.24 &  2.62 & 2.98 & 1.92 & 2.24 & 2.62 & 2.98 \\
 \hline
\end{tabular}
\end{center}
\caption{Trials were drawn from these eight possibilities, two different distances at four different average speeds, constrained by choosing .}
\label{table:protocol}
\end{table}

\begin{figure}[ht!]
\begin{center}
	\includegraphics{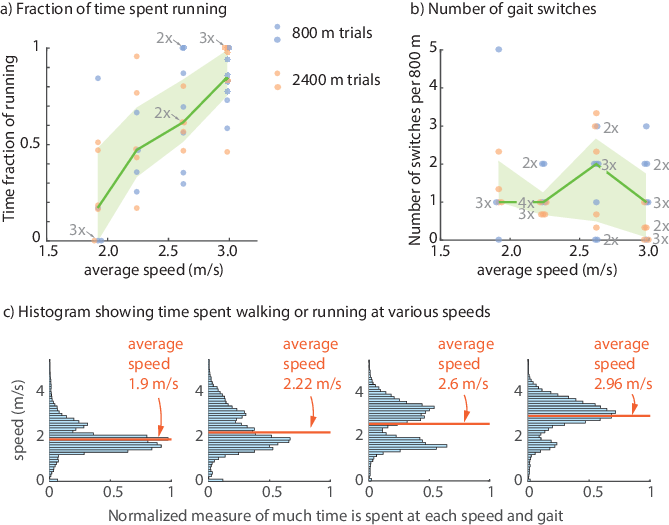}
\caption{\textbf{Humans use a walk-run mixture.} a) Fraction of time spent running for the 800 m trials (blue dots) and the 2400 m trials (orange dots). Fraction of running increases with average speed. The median running fraction (green line) and 25-75$^{th}$ percentile (light green band) are also shown. The labels 2x and 3x refer to the number of multiple coincident data points, also indicated by sideways presentation jitter of the data points, though the prescribed averaged speed is just one of four possibilities indicated in Table \ref{table:protocol}. All 42 data points are represented here. b) Number of switches from running to walking or walking to running. The median number of switches (green line) and 25-75$^{th}$ percentile (light green band) are also shown. Labels 2x, 3x, etc., and sideways presentation jitter of data points added to indicate multiple coincident data points as in panel-a. Three trials in which participants did not reach on time are not plotted. c) Histogram of time spent at various speeds for trials at each of the four average speeds. The histograms pool data over all participants and all trials at the specific average speed. Each histogram has two peaks, one corresponding to walking (low speed) and another corresponding to running (high speed).}
\label{fig:RunningFraction}
\end{center}
\end{figure}

\section{Results}
\paragraph{Humans can achieve average speeds with precision using a clock over long distances.} Participants typically completed the trials in exactly the prescribed amount of time, arriving neither too early nor too late, using walking and/or running, appropriately adjusting their speeds as needed. Only in 3 trials out of 42 trials did the participants travelled so quickly initially that they had to be at virtually zero velocity for more than 10 sec, out of a typical trial duration of at least 540 seconds. This precision of arrival is perhaps to be expected given the countdown clock providing feedback on time remaining, the visible course providing feedback on the distance remaining, and the ability to change speed. 

The participants' GPS-computed average speed was only 0.03 ms$^{-1}$ different from the prescribed speed on average. The GPS-computed average speed and prescribed average speed were fit by the linear regression $v_\mathrm{computed}= 0.99 V_\mathrm{avg} + 0.043$ ($p = 10^{-37}$, $R^2 = 0.991$), which demonstrates human ability to achieve different average speeds with high precision.

\begin{figure}[htbp]
\begin{center}
\includegraphics{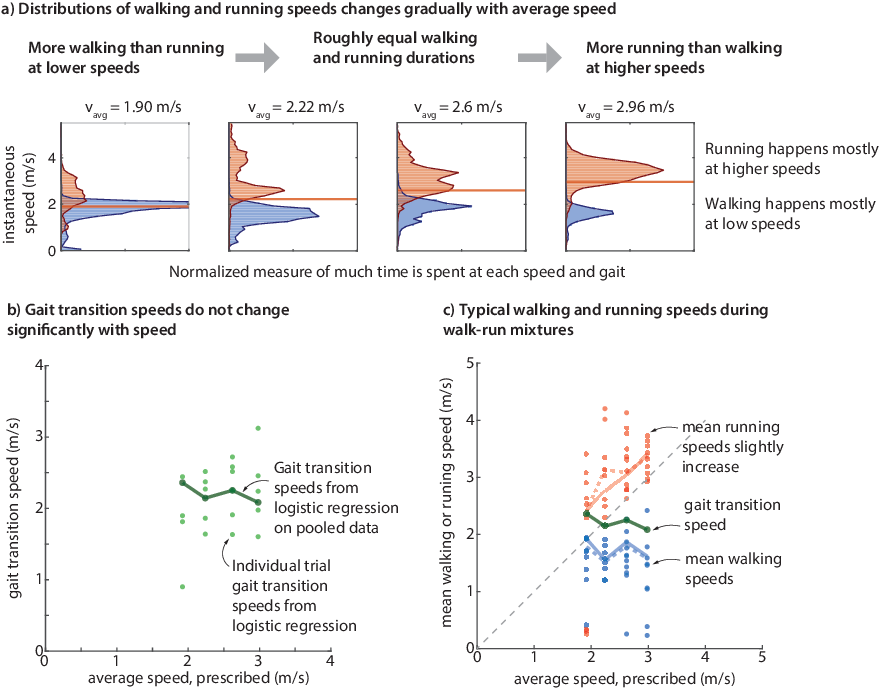}
\caption{a) Distribution of walking and running speeds changes gradually with average speed, with more time spent walking at low speeds and more time spent running at high speeds. Running speeds are systematically higher than walking speeds at each average speed. The overlap observed between walking and running speeds is largely an artifact of pooling participants (see Supplementary Figure \ref{appfig:IndividualSubjects}). b) Gait transition speeds obtained from a logistic classifier do not change significantly speed. c) Mean running speeds increase with average speed. Mean walking speed remains largely constant or very slowly decreases.}
\label{fig:WalkAndRunSpeedTrends}
\end{center}
\end{figure}

\begin{figure}[htbp]
\begin{center}
\includegraphics{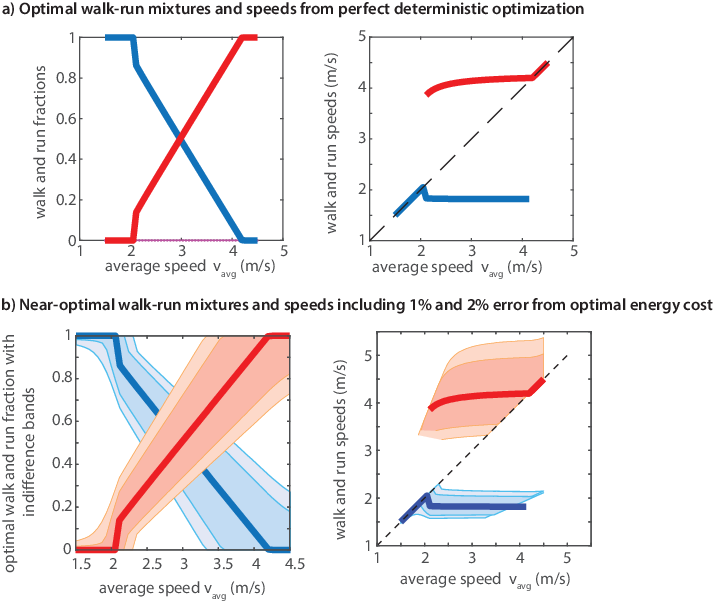}
\caption{a) Optimal walking and running fractions as a function of average speed, assuming the optimization has perfect knowledge and has zero error. During the walk-run mixture regime, the running speed slowly increases and the walking speed remains roughly constant. b) Optimal walk-run fractions and optimal speeds, allowing that the metabolic energy can be within 1\% (darker band) or 2\% (lighter band) of the optimal metabolic cost.}
\label{fig:ModelPredictions}
\end{center}
\end{figure}

\paragraph{Humans mostly used a walk-run mixture for the intermediate average speeds.} Participants used a mixture of walking and running in 90\% of the trials at the two intermediate speeds (2.22 and 2.6 ms$^{-1}$). On average, walking dominates the walk-run mixture at the lower speeds and running dominates the walk-run mixture at the higher speeds (Figure \ref{fig:RunningFraction}a), so that the walk-run mixture gradually changes as speed is increased. The time fraction of walking decreases and the time fraction of running increases with average speed. Thus, for this overground gait transition task, there is no sharp gait transition speed but only a ``gait transition regime'', which has substantial overlap with the speed range 2 m/s to 3 m/s. Despite the substantially greater distances involved in this study (800 m and 2400 m), the overall trends in the walk-run fraction is similar to that observed for trials with much smaller distance (120 m) in the earlier study \cite{LongSrinivasan2013}, suggesting that this walk-run mixture is a natural behavior for this task. 

\paragraph{Walking at low speeds and running at higher speeds.} The speed distributions during each trial have two peaks (Figure \ref{fig:RunningFraction}c). After categorizing every moment as either walking or running, we obtain separate walking speed and running speed histograms for each trial. 
These gait-specific histograms illustrate that in each of the walk-run mixtures, walking generally happens at low speeds and running generally happens at higher speeds (Figure \ref{fig:WalkAndRunSpeedTrends}a; see Supplementary Figure \ref{appfig:IndividualSubjects} for individual trial histograms). Specifically, people accomplish the required average speed by averaging over a slow walking speed and a faster running speed, as qualitatively predicted by energy optimality \cite{LongSrinivasan2013}. As the average speed increases, the speed distribution changes so that there is more time spent at the higher running speeds, resulting in a larger peak at the higher speed: speed 1.98 ms$^{-1}$ is dominated by walking and speed 2.96 ms$^{-1}$ is dominated by running. While the pooled histograms show large gait-specific speed variability (Figure \ref{fig:WalkAndRunSpeedTrends}a), the individual-specific histograms have narrower speed distributions for each gait (Supplementary Figure \ref{appfig:IndividualSubjects}). 

For most participants that used a mixture of walking and running, there is typically no overlap in the speeds used respectively for walking and running: typically there is a substantial gap in the speeds used. This gap is not obvious in the gaitwise histograms pooled across subjects (figure \ref{fig:WalkAndRunSpeedTrends}a) but is clear from visual inspection of the individual subjects' gaitwise speed histograms (appendix figure \ref{appfig:IndividualSubjects}). Thus, subjects do not use intermediate speeds, instead jumping directly from walking speeds to higher running speeds. 

These qualitative features regarding walking speeds and running fractions are generally preserved across participants, even though there is considerable variability in the speeds and the running fractions between participants (Figure \ref{fig:RunningFraction} and \ref{fig:WalkAndRunSpeedTrends}). 

\paragraph{Systematic changes in running speed, but not walking speed or gait transition speed.}
In this gait transition regime, mean running speed increased systematically with average speed, thereby contributing to the increased average speed (figure \ref{fig:WalkAndRunSpeedTrends}b-c). In contrast, mean walking speeds remained roughly constant as average speed increased. We obtained a `within-trial gait transition speed' for each trial by applying a logistic classifier on the walking and running speed data and found that this gait transition speed also does not change systematically with speed, instead remaining roughly constant (figure \ref{fig:WalkAndRunSpeedTrends}b-c).

\paragraph{Predictions from energy optimality agree qualitatively with experiments.} Computational predictions from energy optimality show that it is optimal to use walking alone for locomotion tasks with low average speeds (too much time), it is optimal to use running alone for tasks with high average speeds (very little time), and it is optimal to use a walk-run mixture for intermediate optimal speeds, as observed here (figure \ref{fig:ModelPredictions}a,b). In the walk-run mixture regime, the model predicts that the walking fraction is high for low average speeds and the running fraction is high for high average speeds. Further, the model predicts that in the walk-run mixture regime, the walking speeds remains nearly constant but the running speed increases with average speed (figure \ref{fig:ModelPredictions}a,b), as qualitatively observed in experiments (figure \ref{fig:WalkAndRunSpeedTrends}b-c). The running speed increasing with average speed requires there to be a cost for speed transitions \cite{seethapathi2015metabolic}, as not having any transition costs results in constant walking and running speeds across the different average speeds (see \cite{LongSrinivasan2013} for the version without the transition costs). This result reemphasizes the importance of such transient costs \cite{seethapathi2015metabolic} in predicting human behavior. 

Our sensitivity analysis reveals that the set of speeds and durations within 1\% or 2\% of the optimal energy costs is relatively broad, revealing a flat energy landscape, allowing humans high variability in speed with little energy penalty (figure \ref{fig:ModelPredictions}b-c). This analysis shows that the energy cost is less sensitive to the running speed (larger band of speeds within 1\% or 2\% of optimality) compared to the walking speed; this lower sensitivity is inherited from the lower curvature of the running metabolic rate \cite{Mar76}, so even normal running cost of transport is only weakly sensitive to running speed \cite{Steu09}. Similarly, the near speed independence of running cost per unit distance (low curvature cost landscape) coupled with the curvilinear speed dependence of the walking costs may be responsible for the observed trends in the walking and running speed --- that the running speed increases and the walking speed is relatively constant. 


\paragraph{Humans typically switch gait only a small number of times even for long distances.} 
Pure energy optimality in the absence of fatigue or any uncertainty predicts that in the walk-run mixture regime, there should be exactly one switch between walking and running; multiple switches between walking and running would require more speed and gait transition costs. For three out of the four speeds considered here, the median number of switches is one; speed 2.62 m/s has two switches as median, with some variability around this median (Figure \ref{fig:RunningFraction}b). If we assume the cost of a single switch to be $m\left(V_\mathrm{run}^2-V_\mathrm{walk}^2\right)/2$ approximately \cite{seethapathi2015metabolic,Ush03}, we estimate about 3.4 J per unit body mass when the walking speed is 1.5 m/s and the running speed is 3 m/s. This gait and speed switch cost of 3.4 J/kg is negligible compared to about 2300 J/kg for walking 800 m at 1.5 m/s, using steady walking costs from \cite{Bob60,Sri09}. Despite this cost per switch being negligible, humans only switch gaits a small number of times, still keeping the switching cost negligible.

\section{Discussion}
We have performed overground gait transition experiments over much longer distances than earlier \cite{LongSrinivasan2013} and shown that humans use remarkably similar behavior to short distance bouts. Specifically, we find a gradually morphing walk-to-run transition regime, that involves walk-run mixtures, dominated by walking at lower speeds and by running at higher average speeds. Further, in contrast to treadmill gait transition experiments, here, humans do not use walking up to some instantaneous speed and running above that speed. Instead, within each trial, the overground walk-run mixtures involve transitions from walking at a low speed directly to running at a high speed and vice versa, generally with a gap between these speeds. The avoidance of a range of intermediate speeds has been previously found in free-living animals such as ostriches, emus, horses and gnus \cite{Penn75,Hoyt81,daley2016preferred,watson2011gait}. We might speculate that the hysteretic gait-transition behavior on a treadmill \cite{Die95,Hrel07a,Hrel07b} could be an extrapolation of the overground strategy with such speed gaps, deployed to an unfamiliar treadmill setting. That is, in our overground experiment, the walking speed when humans switch from walking to running is lower than the running speed when humans switch from running to walking, reminiscent of the hysteretic gait transition on a treadmill.

Aside from instantaneous energetics \cite{Mar76,Hoyt81}, gait transition from walking to running has been attributed to muscle force-velocity behavior \cite{Far12}, interlimb coordination variability \cite{kao2003gait}, mechanical load or stress \cite{farley1991mechanical,Die95}, cardiovascular loads \cite{ilic2021does}, and cognitive or perceptual factors such as visual flow \cite{Moh04,guerin2008optical}, all examined on a treadmill; see \cite{kung2018factors} for a review. There is mixed evidence on whether cognitive load influences the walk to run gait transition \cite{daniels2003attentional,voigt2019human}. However, none of these instantaneous factors has been used to explain why there may be hystereses between the walk-to-run and the run-to-walk speeds on treadmills, while Diedrich and Warren \cite{Die95,diedrich1998dynamics} provided a multiple-energy-minima-based explanation of such hysteresis. In contrast, gait transition mechanisms that are dynamical systems and stability-based \cite{voigt2021puzzle,Die95,hansen2017role,raffalt2020walk,gan2018all} cannot immediately explain the natural overground transitions observed here as a function of average speed. Specifically, it is unclear how any of these `instantaneous' theories can explain the gradual walk-run-mixture-based gait transition regime behavior we have demonstrated in this manuscript. Predicting a walk-run mixture over a full bout as being optimal must necessarily require a theory that integrates some performance measure over the entire bout, rather than making decisions instantaneously based on crossing a threshold --- or at the very least, use a control policy that switches gait and speed based on the time and distance remaining. Similarly, within each gait, there are speed fluctuations which may be partly due to feedback control  based on the countdown clock and distance remaining, possibly correcting for any errors  to arrive on time  \cite{tiew2020pre,song2012regulating,thomas2019control}. Such feedback control strategies and any learning effects over multiple trials could be explored in future work. It may also be fruitful to compare the structure of these speed fluctuations and control strategies to those during preferred walking overground walking without a time constraint \cite{collins2013two}, which does have higher instantaneous speed fluctuations than during walking on a treadmill \cite{wang2014stepping}.

Some recent work has also pointed to usefulness of time-like costs or discounting to explain gait-speed choices, in addition to energy, but given such considerations do not apply here due to the precise time durations provided for each locomotor task \cite{shadmehr2016representation,carlisle2023optimization}. One study \cite{summerside2018contributions} included time-related costs to explain the potential preference between walking and running based on distance traveled; in contrast to our study, this study examined fixed gait per bout (allowing no mixtures and encouraging constant speed) and examined explicitly `expressed preference' (what people said they preferred based on trying both choices in a preference experiment) rather than `revealed preference' (what people naturally do under the condition). 

We have considered one kind of overground gait transition, in which the task is traveling a given distance at a desired average sub-maximal speed. Another kind of ecological overground gait transition that might have been common in our evolutionary past is to be walking normally and then having to accelerate to a higher running speed  to either chase prey or evade a predator. Piers et al \cite{pires2014joint} considered such an overground gait-transition task, but theirs is a substantially different task from that considered in this manuscript: from an energy optimality perspective, the task of Piers et al \cite{pires2014joint} will be dominated by the cost of changing speeds and the cost of switching gaits \cite{seethapathi2015metabolic,Ush03} --- so we do not \textit{a priori} expect identical gait transition speeds between their and our experiments. 

The mathematical structure of the gait transition regime with walk-run mixtures is similar to that of phase transitions in materials, for instance, the existence of twinning microstructure (e.g., in an austenite-martensite boundary) or the co-existence of different phases of matter at a given temperature (e.g., water and vapor at room temperature, with the vapor fraction increasing with temperature). In a more complex mathematical model that was able to arbitrarily choose the locomotor patterns  using trajectory optimization \cite{Sri06,srinivasan2007idealized,Sri11}, a walking-like gait was selected at low speeds, a running-like gait was selected at high speeds, and the gait transition phase included a `flat' portion of the energy landscape in which there were infinitely many gaits of equal energy cost in the simple approximation \cite{srinivasan2007idealized}. Our work has focused on a dynamics-free model of gait transition, but future work may involve performing long-timescale gait optimization with a biped model to obtain the optimal gait strategy at each moment \cite{fisher2019intermediate}. Future work could also involve examining what factors mediate the triggering of transitions between walking and running, as well as speed changes within a gait, based on the task requirements. Finally, Long and Srinivasan \cite{LongSrinivasan2013} predicted the occurrence walk-run mixtures in small children (leg length 50-55 cm) walking next to adults, walk-trot mixtures in small dogs on a long leash walking next to a human, stand-walk or stand-run mixtures in small animals or even humans on sufficiently long treadmills, and in human 5-hour marathons, as all of these situations correspond to when such gait mixtures would be energy-optimal. Future studies could test each of these predictions via targeted experiments or observational studies. 

\paragraph{Acknowledgment.}
This work was supported in part by NSF CMMI grant 1254842. We are grateful to Mark Snartese and Max Donelan in suggesting the use of VBOX GPS instead of a less accurate GPS device, enabling the original study \cite{LongSrinivasan2013} as well as this current study.

\paragraph{Competing interests.} We declare that we have no competing interests.

\paragraph{Data and code availability.} Data and code are available publicly at the Zenodo data repository: \\ \url{https://doi.org/10.5281/zenodo.17103631}


\bibliographystyle{vancouver}  
\bibliography{walkRunBakerSrinivasan2024v8NoComments}

\clearpage
\renewcommand{\thefigure}{S\arabic{figure}}
\setcounter{figure}{0} 
\section*{Supplementary Figures}

\begin{figure}[htbp]
\begin{center}
	\includegraphics{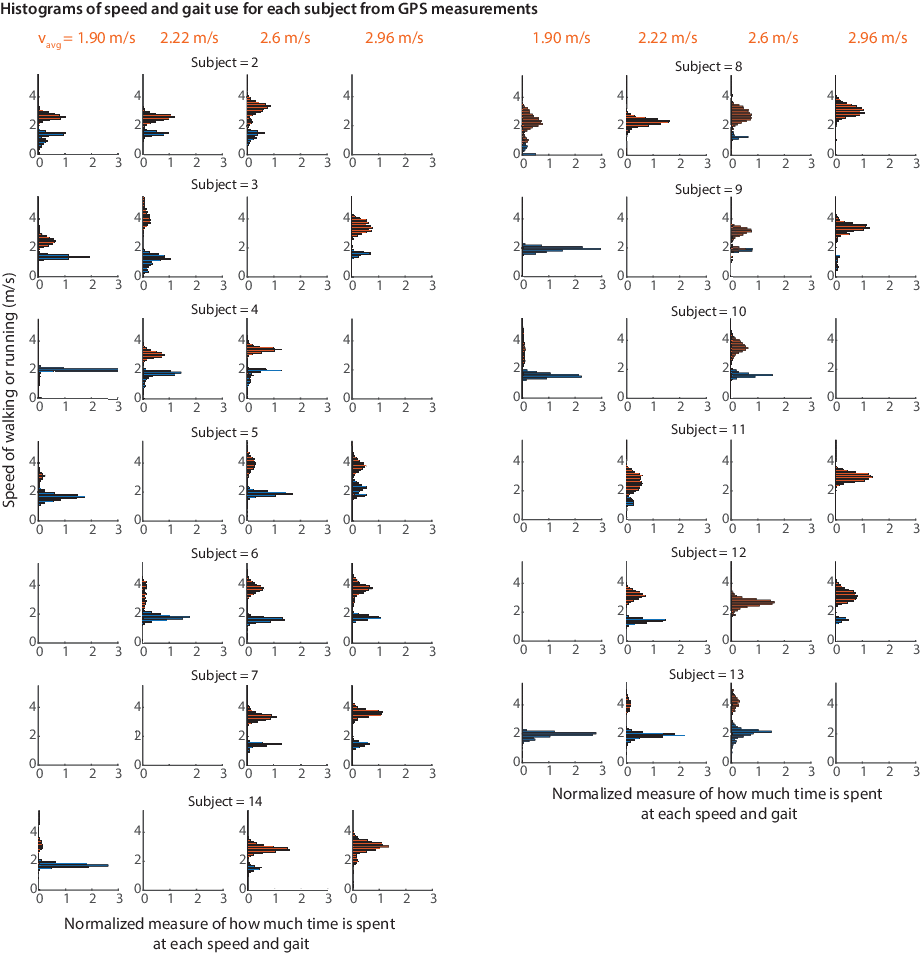}
\caption{The histograms of GPS-derived speeds for each subject classified into walking (blue) and running (red) bouts. This data was pooled across subjects to produce the main manuscript figures \ref{fig:RunningFraction}c and \ref{fig:WalkAndRunSpeedTrends}a. Given the large distances involved, no  subject performed all four average speed tasks, so some axes do not have histogram data.}
\label{appfig:IndividualSubjects}
\end{center}
\end{figure}

\end{document}